\documentclass[aps,prb,twocolumn,footinbibw,superscriptaddress]{revtex4-2}

\usepackage[english]{babel}
\usepackage[utf8]{inputenc}

\usepackage{amsmath}
\usepackage{amssymb}
\usepackage{amsfonts}
\usepackage{bm}
\usepackage{dsfont}

\usepackage{graphicx}

\usepackage{comment}

\usepackage{hyperref}
\hypersetup{
	colorlinks=true,
	linkcolor={cyan!50!black},
	citecolor={red!50!black},
	urlcolor={red!40!black}
	}
 
\usepackage{float}
\graphicspath{{fig_old/}}

\usepackage{xcolor}
\definecolor{myred}{rgb}{0.85,0,0}
\definecolor{mygray}{rgb}{0.87, 0.87, 0.87}

\newcommand{\semi}{{semi-classical} }
\newcommand{\eff}{\mathrm{eff} }
\newcommand{\ext}{\mathrm{ext} }
\newcommand{\qu}{^\mathrm{qu} }
\newcommand{\cl}{^\mathrm{cl} }
\newcommand{\clqu}{^\mathrm{cl/qu} }

\begin{document}

\title{Accounting for Quantum Effects in Atomistic Spin Dynamics}

\author{M. Berritta}
\email{m.berritta@exeter.ac.uk}
\affiliation{Department of Physics and Astronomy, University of Exeter, Stocker Road, Exeter EX4 4QL, UK}
\affiliation{Department of Physics and Astronomy,  Uppsala University, P.\,O.\ Box 516, SE-75120 Uppsala, Sweden}
\author{S. Scali}
\affiliation{Department of Physics and Astronomy, University of Exeter, Stocker Road, Exeter EX4 4QL, UK}
\author{F. Cerisola}
\affiliation{Department of Physics and Astronomy, University of Exeter, Stocker Road, Exeter EX4 4QL, UK}
\affiliation{Department of Engineering Science, University of Oxford, Parks Road, Oxford OX1 3PJ, UK}
\author{J. Anders}
\email{janet@qipc.org}
\affiliation{Department of Physics and Astronomy, University of Exeter, Stocker Road, Exeter EX4 4QL, UK}
\affiliation{Institute of Physics and Astronomy, University of Potsdam, 14476 Potsdam, Germany}


\begin{abstract}

Atomistic spin dynamics (ASD) is a standard tool to model the magnetization dynamics of a variety of materials.
The fundamental dynamical model underlying ASD is entirely classical.
In this paper, we present two approaches to effectively incorporate quantum effects into ASD simulations, thus enhancing their low temperature predictions.
The first allows to simulate the magnetic behavior of a quantum spin system by solving the equations of motion of a classical spin system at an effective temperature relative to the critical temperature. This effective temperature is determined \emph{a priori} from the microscopic properties of the system.
The second approach is based on a \semi model where classical spins interact with an environment with a quantum--like power spectrum. The parameters that characterize this model can be calculated \emph{ab initio} or extracted from experiments.
This \semi model quantitatively reproduces the absolute temperature
behavior of a magnetic system, thus accounting for the quantum mechanical aspects of its dynamics, even at low temperature.
The methods presented here can be readily implemented in current ASD simulations with no additional complexity cost.
\end{abstract}

\maketitle

\section{Introduction}
Magnetic materials modeling is crucial for the understanding of groundbreaking experiments~\cite{bigot96}, as well as for a variety of new technologies~\cite{wolf01,parkin08}. The development of ultra-fast laser pulses led to the discovery of ultra-fast demagnetization induced by thermal fluctuations~\cite{bigot96} and magnetization reversal~\cite{radu11,ostler12}.
These discoveries paved the way for the development of new technologies such as heat assisted magnetic recording~\cite{kryder08}.
Furthermore, thermal effects play a crucial role in helicity-dependent magnetization switching~\cite{lambert14, john17}.

Atomistic spin dynamics (ASD) simulations are among the most effective tools for studying magnetization dynamics in materials with thermal fluctuations~\cite{nowak05,boerner05,evans14,skubic08}.
They model the atoms in the materials as localized classical magnetic moments interacting through an effective Hamiltonian. This Hamiltonian can be fully parameterized with {\em ab initio} calculations and numerically solved, including Markovian thermal fluctuations, to obtain the spin dynamics. 
ASD simulations have been successful in several situations e.g., in predicting the experimentally observed heat-assisted magnetization reversal~\cite{radu11,ostler12}, modeling~\cite{nowak11,chico14} the domain wall dynamics induced by thermal gradients~\cite{franken09,mohrke10} and the current-induced skyrmion dynamics~\cite{fernandes20} observed in~\cite{hrabec17}.
Despite their versatility and predictive capability, ASD simulations, being predominantly classical in nature, exhibit inaccuracies in modeling low-temperature demagnetization driven by quantum excitations like spin waves~\cite{halilov98}.
Accurately accounting for quantum fluctuations would require simulating the full quantum dynamics. The dynamics of coupled quantum spins can be studied with different numerical methods, e.g. quantum Monte Carlo simulations~\cite{kormann11} and tensor-network methods~\cite{alhambra21,kuwahara21}. However, the quantum nature of the problem limits the system size while restricting to high temperatures and short time scales~\cite{xu2022}. 
In fact, given the current hardware limitations, the dynamics of more than a few dozens of quantum spins is already inaccessible. In contrast, one of the appealing characteristics of ASD simulations is their capability to model the dynamics of millions of coupled classical spins~\cite{evans14,skubic08,ma08}. This makes them suitable for studying systems where the translational symmetries are broken, e.g. by defects or local perturbations.

To improve the accuracy of ASD simulations in the low temperature regime, in Ref.~\cite{evans15} the authors employ a heuristic approach to reproduce the experimental temperature dependence of a material's magnetization. 
In this approach, they postulate that the experimental temperature $T_{\rm exp}$ and the atomistic spin dynamics simulation temperature $T_{\rm sim}$ are related by a power-law, as $\tau_{\rm sim}=\tau_{\rm exp}^\alpha$, where $\tau_{\rm sim/exp}=T_{\rm sim/exp}/T_c^{\rm sim/exp}$ and $T_c^{\rm sim/exp}$ is the Curie temperature of the material.
The required rescaling exponent $\alpha$ was obtained by comparing the results of the ASD simulations with experimental measurements.
Importantly the temperature rescaling not only reproduces the systems equilibrium, but it also 
captures the behavior of the ultra-fast magnetization dynamics in pump-probe experiments~\cite{bigot96}. While this approach seems effective across various situations, it is purely heuristic, lacking a microscopic understanding of its underlying mechanisms.

An alternative approach to account for quantum effects in ASD simulations was proposed in Ref.~\cite{barker19}, where the authors solve (Markovian) Landau-Lifshitz-Gilbert dynamics, but include stochastic noise in the spin dynamics which has a quantum power spectrum. Thanks to this approach, it is possible to reliably model the low-temperature behavior of the magnetization of materials. However, the high temperature behavior was not fully captured, i.e.  Curie temperatures are somewhat overestimated.

Building on these previous works, in this paper we first provide an analytical expression and a physical explanation for the power-law rescaling of Ref.~\cite{evans15}. This is done  in Sec.~\ref{sec:temp_res}, where we compare the classical spin model to its quantum mechanical counterpart in the mean field framework. By means of this comparison, we derive rescaling power-laws, which turn out to be in good agreement with the ones found in Ref.~\cite{evans15}.
Strikingly, we do not need preliminary ASD simulations to calibrate the rescaling exponent.
Our predictive method can be readily applied to ASD simulations to account for dynamical quantum effects at no additional computational cost.

Secondly, in Sec.~\ref{sec:open-heisenberg} of this paper we follow the quantum noise spectrum approach of \cite{barker19} and improve it to better capture the high temperature regime and  predict critical temperatures. To achieve this we directly solve the steady-state of a classical spin system interacting with a quantized non-Markovian bath. This approach does not require any rescaling and quantitatively reproduces the low temperature behavior of the magnetization as well as its critical temperature $T_c$. This model can be fully parameterized by means of {\em ab initio} calculations or experiments and constitutes a highly promising candidate for improving the accuracy of ASD simulations by accounting for quantum effects.


Before reporting on these two methods to include quantum effects in ASD, we here introduce the general setting.

\medskip


\section{General setting}
We consider the well-known Heisenberg model~\cite{auerbach12},  which is described by the Hamiltonian 
\begin{equation}
\label{eq:heisenberg}
\mathcal{H}_{\rm H}= - \frac{J}{2} \sum_{\langle i,j\rangle}\hat{\bm{S}}^{(i)} \cdot \hat{\bm{S}}^{(j)} - \gamma H_\ext \sum_j \hat{S}^{(j)}_z
\end{equation}
where $J$ is the exchange parameter which can be calculated {\em ab initio}~\cite{liechtenstein}, $\gamma$ is the electron gyromagnetic ratio, $H_\ext$ is an external magnetic field pointing in the $z$ direction, the sum over $i$ runs over all the lattice sites, and $\langle i,j\rangle$ over nearest-neighbor pairs. For a periodic system made of spins of length $S_0=n\hbar/2$ with $n$ a natural number, the mean field approximation (MFA) self-consistent equation for the expectation value of
the normalized spins~\footnote{Note that the normalized spin can also be expressed as $s_z = M(T)/M_0$ where $M(T)$ is the magnetization at temperature $T$ and $M_0$ is the magnetization at $T=0$.} $s_z=\langle S_z\rangle/S_0$ is ${s_z\clqu=\mathcal{F}\clqu\left(\beta\gamma H_\eff(s_z)S_0\right)}$~\cite{solyom07}.
Here, $\beta=1/k_BT$ with $k_B$ the Boltzmann constant and $T$ the temperature,
$\mathcal{F}\clqu(x)$ is the Langevin function $\mathcal{F}\cl(x)=\coth(x)-1/x$ for the classical Heisenberg model, and the Brillouin function
$ \mathcal{F}\qu(x)=\{(n+1)\coth\left[(n+1)x/n\right]-\coth\left(x/n\right)\}/n$ 
for the quantum Heisenberg model with $n=2S_0/\hbar$~\cite{solyom07}.
Furthermore, ${H_\eff(s_z)=H_{\ext}+\mathsf{z} JS_0s_z/\gamma}$ is the effective magnetic field felt by each spin in the lattice, with $\mathsf{z}$ being the coordination number. 

It is insightful to express the magnetization curves $s_z\clqu$ as a function of the relative temperatures ${\tau^\mathrm{cl/qu} = T/T_\mathrm{c}^\mathrm{cl/qu}}$. Here, $T_\mathrm{c}^\mathrm{cl/qu}$ are the critical temperatures marking the transition from the trivial solution $s_z\clqu=0$ to the non-trivial solutions $s_z\clqu\neq0$.
In the absence of an external magnetic field, $H_\ext=0$, the critical temperatures are
$T_\mathrm{c}\clqu = \mathsf{z}J\langle S^2\rangle\clqu / (3k_\mathrm{B})$,
where $\langle S^2\rangle\cl = S_0^2$ for the classical case and $\langle S^2\rangle\qu = S_0(S_0+\hbar)$ for the quantum one. These critical temperatures are typically overestimated since the MFA does not account for the system's fluctuations~\cite{evans14}. Note that, in terms of the relative temperatures $\tau\clqu$, the expectation value of the normalized spin takes the self-consistent form
\begin{equation}
    \label{eq:self_consistent_simple}
    s_z\clqu = \mathcal{F}\clqu\left(\frac{3}{\tau\clqu} \frac{S_0^2}{\langle S^2\rangle\clqu} s_z\clqu\right).
\end{equation}
Therefore, in the MFA for the Heisenberg model the dependence of $s_z\clqu$  is reduced to two free parameters, namely the relative temperature $\tau\clqu$ and the spin length $S_0$.

\begin{figure}
    \centering
    \includegraphics{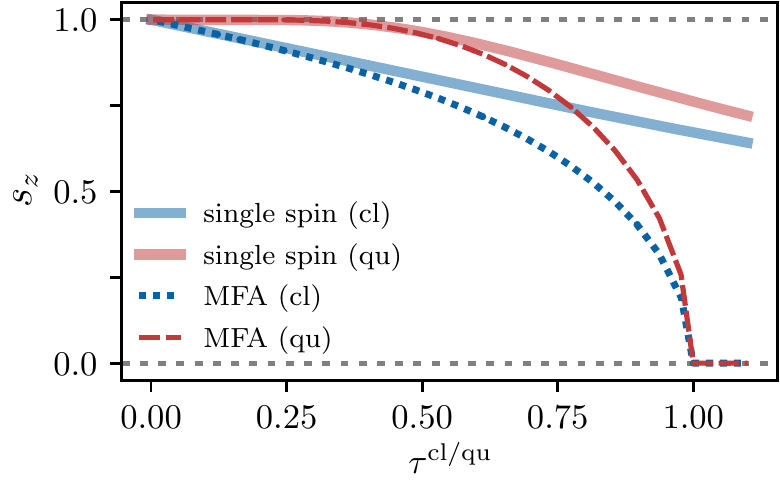}
    \caption{\textbf{Quantum and classical MFA magnetization}. We show the normalized spin magnetization $s_z$ for a spin system with $S_0=\hbar/2$ (see Eq.~\eqref{eq:self_consistent_simple}) evaluated with the MFA as function of the relative temperature $\tau\clqu=T/T_c\clqu$ for classical spin vectors (blue dotted line) and quantum spin operators (red dashed line). Note that the only effect of $J$ and $\mathsf{z}$ is to determine $T_c$. Since here we calculate the magnetization as a function of the relative temperature $\tau^{\rm cl/qu}=T/T_c^{\rm cl/qu}$, $J$ and $\mathsf{z}$ do not need to be specified. 
    We also plot the classical (blue solid line) and quantum (red solid line) single-spin $s_z$ curves in a field $H_\eff^0=\mathsf{z}JS_0/\gamma$ .}
    \label{fig:cl_vs_qu}
\end{figure}

The deviation of classical ASD simulations from experiments is most prominent at low temperatures~\cite{evans15}, particularly so for magnetic materials with atoms carrying a small magnetic moment. Indeed, it is in this regime that quantum contributions play an important role in spin systems causing discrepancies between the classical and quantum results~\cite{fisher64,millard71,cerisola22}. To see how these deviations arise, we can consider the spins at low temperature to be approximately in the ground state. The spins then fully align along the $z$ direction ($s_z\simeq1$) and the effective field becomes independent of $s_z$, $H_\eff^{T\simeq0}\simeq\mathsf{z}JS_0/\gamma$. Thus, in this limit, the dependence of the right hand side of Eq.~\eqref{eq:self_consistent_simple} on $s_z$ can be neglected. This makes the expectation value of each spin independent of the expectation values of the other spins in the system, reducing the multi-spin problem to a single-spin problem in the low-$T$ limit (see Fig.~\eqref{fig:cl_vs_qu}).
While a single classical spin has a continuous energy spectrum, the quantum counterpart does not. This implies that, already at the smallest finite temperature, the classical spin can experience infinitesimally excited states such that the expectation value $s_z$ deviates from the ground state $s_z=1$. On the other hand, the expectation value of the quantum spin remains in the ground state $s_z=1$ as long as the thermal energy $k_\text{B}T$ is smaller than the energy gap between the ground state and the first excited state, $\Delta E=\gamma H_\eff^{T=0}\hbar$.

In Fig.~\ref{fig:cl_vs_qu}, we show the comparisons between the expectation value of a single-spin in an effective field $H_\eff^0$ and the expectation value of a multi-spin system obtained with the MFA as functions of the relative temperature $\tau\clqu$. This comparison is shown for both quantum and classical cases. In the limit of zero temperature, the single-spin calculations (solid lines) closely resemble those of the multi-spin case
obtained by means of the MFA (dotted and dashed lines). Note also that, in both the single-spin as the multi-spin case, while the classical spin (cyan solid and dotted lines) has a finite gradient at $T=0$, the quantum spin (red solid and dashed lines) is characterized by a plateau which only drops once $\tau\qu \gtrsim 0.4$. That is, the classical spin is susceptible to arbitrarily small temperatures while, for the quantum case, there is a minimum amount of thermal energy needed to excite the system. 

We conclude that the difference between the quantum and classical MFA can be tracked down to the single-spin qualitative behaviour in an external field. This observation is key in designing the approach described in Sec.~\ref{sec:open-heisenberg}. There the assumption is used that it is possible to account for quantum effects by just coupling each individual classical spin of the system to an environment with a quantum--like power spectrum.


\begin{figure*}
    \centering
    \includegraphics{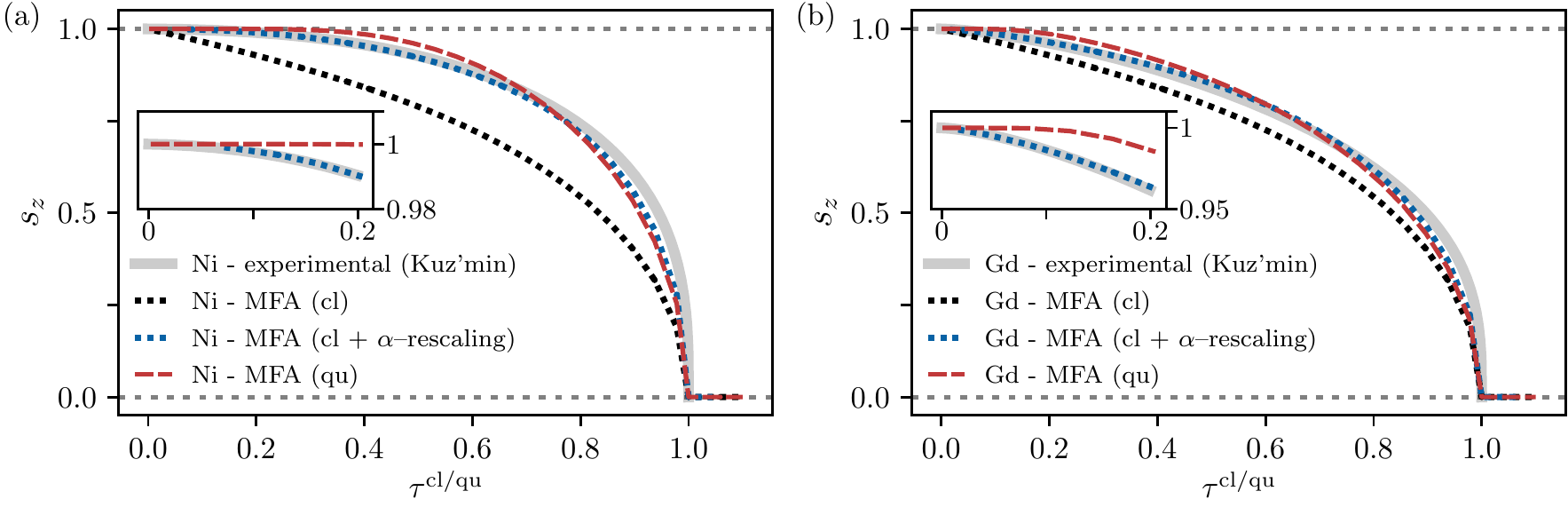}
    \caption{\textbf{Temperature-rescaled equilibrium magnetization for nickel and gadolinium}. We show the comparison between the results for Ni (left panel) and Gd (right panel) of the classical MFA $s_z^{\rm cl}$ with a temperature rescaling according to Eq.~\eqref{eq:Tscaling} (blue dotted lines) and the quantum MFA $s_z^{\rm qu}$ (red dashed lines). Also shown are the phenomenological Kuz'min curves (gray solid lines).
    Additionally, the classical MFA predictions are shown (black dotted lines) to highlight their discrepancy to the experiments.
    The insets zoom into the low temperature behavior of the same curves.}
    \label{fig:panels}
\end{figure*}

\section{Power law temperature rescaling in the MFA}
\label{sec:temp_res}
The experimental behaviour of the magnetization at low temperature is known to follow the Bloch's law $M(T)\simeq M_0(1-(\tau^{\rm exp})^{3/2})$. On the other hand, classical ASD simulations, in the same temperature range, are known to produce a linear magnetization dependence on the temperature $M(T)\simeq M_0(1-\gamma\tau^{\rm sim})$.
To overcome this discrepancy, in Ref.~\cite{evans15}, Evans {\em et. al.} introduced a heuristic temperature rescaling to ASD simulations to successfully reproduce the experimental temperature dependence of the magnetization. This method proves effective in several contexts and it is now a staple in the ASD community's toolkit. Nevertheless, it lacks a detailed physical interpretation which we will provide below. 
This temperature mapping from the classical to the quantum behavior was done by means of the rescaling
\begin{equation} \label{eq:Tscaling}
    \tau^{\rm sim} = (\tau^{\rm exp})^\alpha,
\end{equation}
for a power $\alpha\in\mathds{R}$ that was fixed by fitting the simulations to
experimental data.
The estimation of $\alpha$ required the comparison of classical ASD simulations with the experimental $M(T)$ curve or alternatively the Kuz'min law. The latter is a phenomenological equation able to accurately reproduce the experimental $M(T)$ while also complying with the Bloch's law~\cite{kuzmin05}.
This same rescaling was found successfully matching the experimentally observed ultra-fast demagnetization in nickel~\cite{bigot96,evans15}. Despite the successes of this method two main questions persist: what is the microscopic origin of such rescaling? Is there a strategy that allows to calculate the rescaling exponents that does not require prior knowledge of experimental results?

Here, we provide a physical interpretation for the mechanism behind the success of such rescaling. We propose a method for calculating the rescaling exponent $\alpha$ without recurring to any prior experimental knowledge about the material's behaviour. Additionally, we give an analytical expression to evaluate $\alpha$ suggesting its sole dependence on the spin length $S_0$.
Our method can be readily used in ASD simulations to improve the low-temperature description without additional computational cost. 

To understand the origin of this rescaling, we first note that the quantum-classical difference is already apparent in the MFA expression~\eqref{eq:self_consistent_simple}. The key idea in this section, is to fit the curves of the classical MFA solution $s_z\cl$ to the curves of the quantum MFA solution $s_z\qu$, assuming a temperature rescaling of the type shown in Eq.~\eqref{eq:Tscaling}, that is $\tau\cl=(\tau\qu)^\alpha$ for various exponents $\alpha$. When expressing the temperature in terms of $\tau = T/T_\mathrm{c}$, the
only free parameter left in Eq.~\eqref{eq:self_consistent_simple} is the spin length $S_0$.
Therefore, to find the 
best fitting exponent $\alpha$ for a given material will
require an appropriate choice of $S_0$.
To do so, we write $S_0 = n\hbar/2$ with $n=1,2,\dots$, and note that a spin
of length $\hbar/2$ carries one Bohr magneton. We then choose the smallest integer $n$ that obeys $n \geq \mu[\mu_\mathrm{B}]$, where
$\mu[\mu_\mathrm{B}]$ is the atomic magnetic moment of the material expressed
in units of Bohr magnetons.
These magnetic moments are obtained e.g. by means of the augmented spherical
wave (ASW) implementation of density functional theory~\cite{eyert12}.
Finally, we determine $\alpha$ in Eq.~\eqref{eq:Tscaling} by minimizing
the mean squared error
\begin{equation}
\label{eq:opt}
    \underset{\alpha}{\operatorname{argmin}} 
    \int_0^1\mathrm{d}\tau\left[s_z\qu(\tau) - s_z\cl(\tau^\alpha)\right]^2,
\end{equation}
where $s_z\clqu$ are evaluated using Eq.~\eqref{eq:self_consistent_simple}.

The temperature rescaling exponent determined by this MFA based method is applicable within a specific temperature range where the relevant microscopic characteristics of the materials remain constant. 
For instance, in the case of nickel and gadolinium, this method is expected to be effective within their temperature ranges  $[0, T_c]$. However, for materials like cobalt, which exhibit a coexistence of hcp and fcc phases, it becomes challenging to uniquely determine the coordination number $\mathsf{z}$. Indeed we expect our approach for calculating the rescaling exponent to work as long as the material in question has a single coordination number $\mathsf{z}$ across the entire temperature range. Additionally, when the Heisenberg model proves inadequate due to factors such as the presence of magnetic anisotropy, as in the case of iron, this method may require extensions.

To illustrate the results obtained with the procedure discussed above, we look at the case of Ni and Gd
whose atomic magnetic moments are $\mu_{\mathrm{Ni}} = 0.66\mu_{\mathrm{B}}$ and $\mu_{\mathrm{Gd}}=7.63\mu_{\mathrm{B}}$ respectively, translating into a spin length $S_0^{\mathrm{Ni}} = \hbar/2$ and $S_0^{\mathrm{Gd}}=8\hbar/2$. 
The coordination numbers for both Ni and Gd are $\mathsf{z}=12$, which are constant in $[0,T_c]$. (Note that the numerical value isn't practically needed for calculations in terms of the relative temperature, see Eq.~\eqref{eq:self_consistent_simple}.)

In Fig.~\ref{fig:panels}, we compare the magnetization curves for Ni and Gd: $s_z\qu(\tau\qu)$ predicted by the
quantum Heisenberg model, the classical prediction with rescaled temperature $s_z\cl((\tau\qu)^\alpha)$, and the Kuz'min phenomenological model~\cite{kuzmin05} which closely matches experiments.
We further plot the bare (without rescaling) classical prediction $s_z\cl(\tau\cl)$. 
As expected from the previous discussion, the standard classical and quantum MFA calculations disagree, particularly at low temperatures. This discrepancy is notably more pronounced for materials with low $S_0$ values, such as Ni, compared to those with higher spin values like Gd. The higher spin in Gd leads to behavior that more closely aligns with classical expectations~\cite{cerisola22}.
Crucially, comparing in Fig.~\ref{fig:panels} the quantum MFA and the classical temperature-rescaled magnetizaion we observe a strong agreement between the two, particularly at low temperatures.
We further quantify the difference between these two predictions by the global relative mean square error,
$\sqrt{\frac
    {\int_0^1\mathrm{d}\tau\left[s_z\qu(\tau) - s_z\cl(\tau^\alpha)\right]^2}
    {\int_0^1\mathrm{d}\tau\left[s_z\qu(\tau)\right]^2}}
$, and find that it is at most 2.5\%, and decreases as $S_0$ increases as expected \cite{cerisola22}.
We further note that in the low temperature regime the agreement between the rescaled classical magnetization and the experimental data is even more pronounced. This is likely due to the fact that the temperature rescaling adheres to a power law, which accurately reproduces the Bloch's law, 
a feature not captured by the bare quantum MFA due to the fact that the bare MFA does not accurately account for fluctuations. 



\begin{figure}[t]
    \centering
    \includegraphics{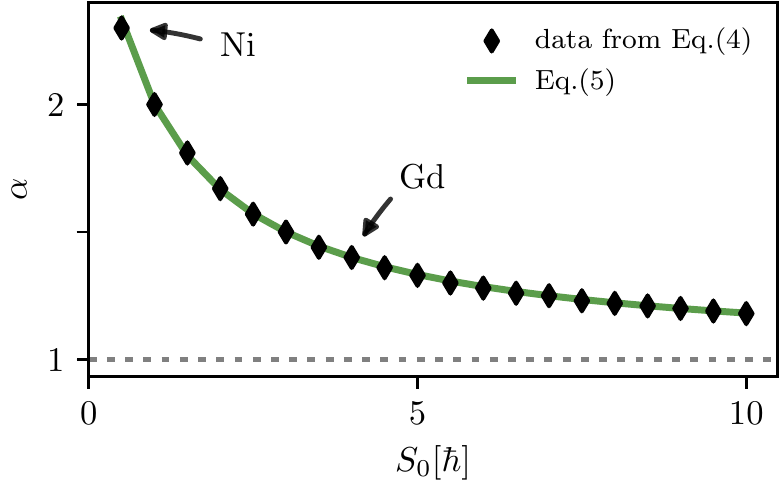}
    \caption{\textbf{Rescaling power $\alpha$ as a function of spin length $S_0$}. Here we plot the dependence of the rescaling parameter $\alpha$, see Eq.~\eqref{eq:Tscaling}, obtained by minimizing Eq.~\eqref{eq:opt}, as a function of spin length $S_0$ (black diamonds). An analytical function,  Eq.~\eqref{eq:alpha_S0_fit}, that very closely matches the obtained exponents is also plotted (green line).}
    \label{fig:exponents}
\end{figure}

Repeating the procedure outlined by Eq.~\eqref{eq:opt}, we obtain the different
scaling exponents $\alpha$ as a function of $S_0$. The values for different $S_0$ are shown in Fig.~\ref{fig:exponents} (black diamonds). Remarkably, we find that the values obtained by this numerical minization procedure can be described to very high accuracy by the simple function (green solid line in Fig.~\ref{fig:exponents})
\begin{equation} \label{eq:alpha_S0_fit}
    \alpha(S_0) = 1 + \frac{2}{1 + S_0/\hbar}.
\end{equation}
Indeed, this extremely simple expression matches the rescaling exponent $\alpha(S_0)$ from Eq.~\eqref{eq:opt} with less than 0.05\% error.

Remarkably, our derived rescaling exponent $\alpha$, predicts a value very similar to the one used in Ref.~\cite{evans15} for nickel (we obtain $\alpha_{\rm Ni}=2.31$ vs $\alpha_{\rm Ni}=2.32$ found by the authors in Ref.~\cite{evans15}) and gadolinium (we obtain $\alpha_{\rm Gd}=1.38$ vs $\alpha_{\rm Gd}=1.28$ found by the authors in Ref.~\cite{evans15}).
In Ref.~\cite{evans15}, finding the exponent $\alpha$ required
running ASD simulation for a wide range of temperatures
and fitting the results to experimental data.
Here, instead, we determined $\alpha$ (Eq.~\eqref{eq:alpha_S0_fit}) by matching the quantum and
classical Heisenberg mean-field predictions.
Using the analytical expression in Eq.~\eqref{eq:alpha_S0_fit}, we can find $\alpha$
for a specific material in a self--contained fashion, that is, eliminating the need for fitting with experimental data.
Including the rescaling described by Eq.~\eqref{eq:alpha_S0_fit} in ASD simulations, one can theoretically predict equilibrium magnetization and
magnetization dynamics without previous reliance on experimental data.

In conclusion, to account for quantum contributions at a given relative experimental
temperature ${\tau_{\mathrm{exp}} = T_{\mathrm{exp}}/T_{\mathrm{c}}^\mathrm{exp}}$,
one can use standard ASD simulations at a temperature
$\tau_{\mathrm{sim}} = (\tau_{\mathrm{exp}})^{\alpha(S_0)}$,
where $\alpha(S_0)$ is fully determined by Eq.~\eqref{eq:alpha_S0_fit}.

\section{Accounting for environmental fluctuations}
\label{sec:open-heisenberg}
The temperature rescaling strategy just discussed is based on the relative temperature $\tau$. Therefore, to predict the magnetization as a function of temperature with real units, it requires prior knowledge of the experimental critical temperature $T_\mathrm{c}^\mathrm{exp}$ of the material.
Here, we propose a second approach, that predicts the magnetization against the absolute temperature with no need for prior knowledge of $T_c^{\rm exp}$. This new method is fundamentally different from the rescaling discussed in Sec.~\ref{sec:temp_res}, and is based on a physically motivated microscopic model where the spins are additionally strongly coupled to a quantum environment. This model can further be fully parameterized through {\em ab initio} calculations. 
As we will see, this approach accurately reproduces the temperature dependence of nickel's magnetization, including low-temperature behavior and a close approximation of the experimental critical temperature.\\
In studying magnetic materials, it is essential to consider the interaction between atomic magnetic moments (spins) and non-magnetic degrees of freedom like phonons. To incorporate these degrees of freedom, we linearly couple the spins in the Heisenberg model $\mathcal{H}_H$ (described by Eq.~\eqref{eq:heisenberg}) to independent bosonic baths as in Ref.~\cite{anders20},
\begin{equation}
\label{eq:open}
    \mathcal{H}_{\rm H+B}=\mathcal{H}_{\rm H}+\mathcal{H}_{\rm I}+\mathcal{H}_{\rm B} ,
\end{equation}
where $\mathcal{H}_{\rm B}=\frac{1}{2}\sum_j\int_0^\infty d\omega\left(\left(\bm{\Pi}_\omega^{(j)}\right)^2+\omega^2\left(\bm{X}_\omega^{(j)}\right)^2\right)$ describes a continuous frequency reservoir at each lattice site $j$ with $\bm{X}_\omega^{(j)}$ and $\bm{\Pi}_\omega^{(j)}$ being the positions and momentum operators of the 3D reservoir oscillators with frequency $\omega$. The interaction term between the magnetic system and the reservoir is 
\begin{equation}
    \mathcal{H}_{\rm I}=-\gamma\sum_j \bm{S}^{(j)}\cdot\int_0^\infty d\omega \, \mathcal{C}_\omega \, \bm{X}_\omega^{(j)},    
\end{equation}
where $\mathcal{C}_\omega$ determines the coupling between the reservoir oscillators and the spins $S^{(j)}$ at lattice position $j$.
This approach serves as a microscopic model for the Landau-Lifshitz-Gilbert (LLG) equation, including inertial terms~\cite{anders20}. We assume isotropic coupling of the spins to these external modes i.e. $C_\omega$ is a scalar, not a tensor~\cite{hartmann23}, and that the spin-phonon coupling $C_\omega$ is the same for all spins (i.e. it does not depend on the lattice index $j$).
From the theory of open systems, the effects of the reservoir are summarized by the spectral density $\mathcal{J}(\omega)=\mathcal{C}_\omega^2 / (2\omega)$.
The spectral density has a well-defined microscopic meaning and, in the case of phonons, is directly linked to the phonon density of state~\cite{nemati22}. The phonon density of state can further be obtained from first principle calculations~\cite{phonopy}. For our simulations on nickel, we parameterize the spectral density as a Lorentzian ${\mathcal{J}(\omega) = (\eta/\pi)\omega_0^4\;\omega/((\omega^2-\omega_0^2)^2 + \omega^2\Gamma^2)}$, which we fit to the experimental phonon density of states reported in Ref.~\cite{kresch07}. This sets a peak frequency at $\omega_0=6$~THz and its broadening at $\Gamma=8$~THz. The coupling strength is set by the dimension-free Gilbert damping $\eta$~\cite{anders20}, taken here as $\eta=0.01$, in line with values typically reported in the literature~\cite{walowski08,evans14}.

The full quantum treatment of the Heisenberg model is unfeasible~\cite{stinchcombe67}, even more so in the case of an Heisenberg model strongly coupled to a bosonic bath. Our main aim here is to develop a method that is capable to capture relevant quantum effects of the magnetization dynamics and equilibrium properties, without dramatically increasing the computational complexity of standard spin vectors ASD methods. Therefore, in line with standard ASD models, we will assume that the spins are classical and furthermore they interact with a bath of harmonic oscillators. Key, however, is that we will impose the environment oscillators obey a quantum-like power spectrum~\cite{barker19,halilov98}. This is reminiscent of the techniques employed in Ref.~\cite{schmid82} for the exact modeling of quantum Brownian motion via a semi-classical model. In Ref.~\cite{barker19}, a quantum environmental power spectrum 
\begin{equation}
    \label{eq:semicl}
    P_\mathrm{sc}=\left(\coth\left(\frac{\hbar\omega}{2k_\mathrm{B}T}\right)-1\right)\mathcal{J}(\omega)
\end{equation} 
with $\mathcal{J}(\omega)= \eta _G  \hbar \omega$, where $\eta_G$ is the Gilbert damping, was used, which eliminates zero point fluctuations of the environment.
Here, we employ the same strategy but where $\mathcal{J}(\omega)$ is now the Lorentzian parametrized for Ni as described above.

To study the effects of this quantum environment on the equilibrium properties of the magnetization, we extend the MFA to account for the interaction with the environment (EMFA).
We perform the usual decomposition of the spins as $\mathbf{S}^{(j)}=\langle \mathbf{S}^{(j)}\rangle+(\mathbf{S}^{(j)}-\langle \mathbf{S}^{(j)}\rangle)$, where the first term represent the ferromagnetic order and the second term captures the fluctuations.
Inserting this expression in the total Hamiltonian of Eq.~\eqref{eq:open} and neglecting the quadratic terms in the fluctuations we obtain
\begin{equation}
    \label{eq:first_or_fl}
    \mathcal{H}_{\rm H+B}
    \simeq\frac{J}{2}\sum_{\langle i,j\rangle}\langle \mathbf{S}^{(i)}\rangle\cdot\langle\mathbf{S}^{(j)}\rangle+\mathcal{H}_B-\gamma\sum_j\mathbf{H}^{(j)}_\mathrm{EMFA}\cdot\mathbf{S}^{(j)},
\end{equation}
where the effective magnetic field generated by the spin $\mathbf{S}^{(j)}$'s neighbor and the harmonic oscillator bath is $\mathbf{H}^{(j)}_{\rm EMFA} =\frac{J}{\gamma}\sum_{i (j)} \langle\mathbf{S}^{(i)}\rangle +\int_0^\infty d\omega \, C_\omega \, \hat{\mathbf{X}}^{(j)}_\omega$ where the sum is over the $i$ neighbours of $j$. The Gibbs state of the total system is $\rho=e^{-\beta\mathcal{H}_{\rm H+B}}/{\rm Tr}[e^{-\beta\mathcal{H}_{\rm H+B}}]$.
The first term in Eq.~\eqref{eq:first_or_fl} is constant and cancels in the Gibbs state. Hence the expectation value of the spin $\mathbf{S}^{(j)}$ can now be calculated using the Gibbs distribution as 
\begin{equation}
    \langle\mathbf{S}^{(j)}\rangle=\frac{{\rm Tr}[\bm{S}^{(j)}e^{-\beta(\mathcal{H}_B-\gamma\bm{H}^{(j)}_{\rm EMFA}\cdot\bm{S}^{(j)})}]}{{\rm Tr}[e^{-\beta(\mathcal{H}_B-\gamma\bm{H}^{(j)}_{\rm EMFA}\cdot\bm{S}^{(j)})}]}
\end{equation}
where the trace ${\rm Tr}[\circ]$ is performed over the bath and the spin degrees of freedom.
Assuming that all spins behave in the same way, i.e. $\mathbf{S}^{(j)}= \mathbf{s} \, S_0$ for all $j$, 
we obtain the quantum phononic Environment+spin neighbours Mean Field Approximation (EMFA) equations for the case of an open Heisenberg model as
\begin{eqnarray}
    \label{eq:os--mfa}
    \langle \mathbf{s}\rangle
    &=&\frac{\text{Tr}[\mathbf{s} \, e^{-\beta(H_\mathrm{B}-\gamma \mathbf{H}_{\rm EMFA}[\langle \mathbf{s} \rangle] \cdot \mathbf{s} \, S_0)}]}{\text{Tr}[e^{-\beta(H_\mathrm{B}- \gamma \mathbf{H}_{\rm EMFA}[\langle \mathbf{s} \rangle] \cdot \mathbf{s} \, S_0)}]} \nonumber \\
     &\equiv&\mathcal{\bm F}_\mathrm{EMFA}(\mathbf{H}_{\rm EMFA}[\langle \mathbf{s}\rangle]/k_B T),
\end{eqnarray}
where $\beta = 1/k_B T$ and
\begin{equation}
    \label{eq:eff_h} \mathbf{H}_{\rm EMFA}[\langle \mathbf{s} \rangle]=\frac{J}{\gamma} \mathsf{z} \, S_0 \, \langle \mathbf{s} \rangle +\int_0^\infty d\omega  \, C_\omega \hat{\mathbf{X}}_\omega.
\end{equation}
Eq.~\eqref{eq:os--mfa} defines a self-consistent equation for the expectation value of the spins in the system. 

In summary, to solve the EMFA equation (Eq.~\eqref{eq:os--mfa}) we proceed as follows (see Fig.~\ref{fig:flowchart}):

\begin{figure}
    \centering
    \includegraphics[width=0.92\linewidth]{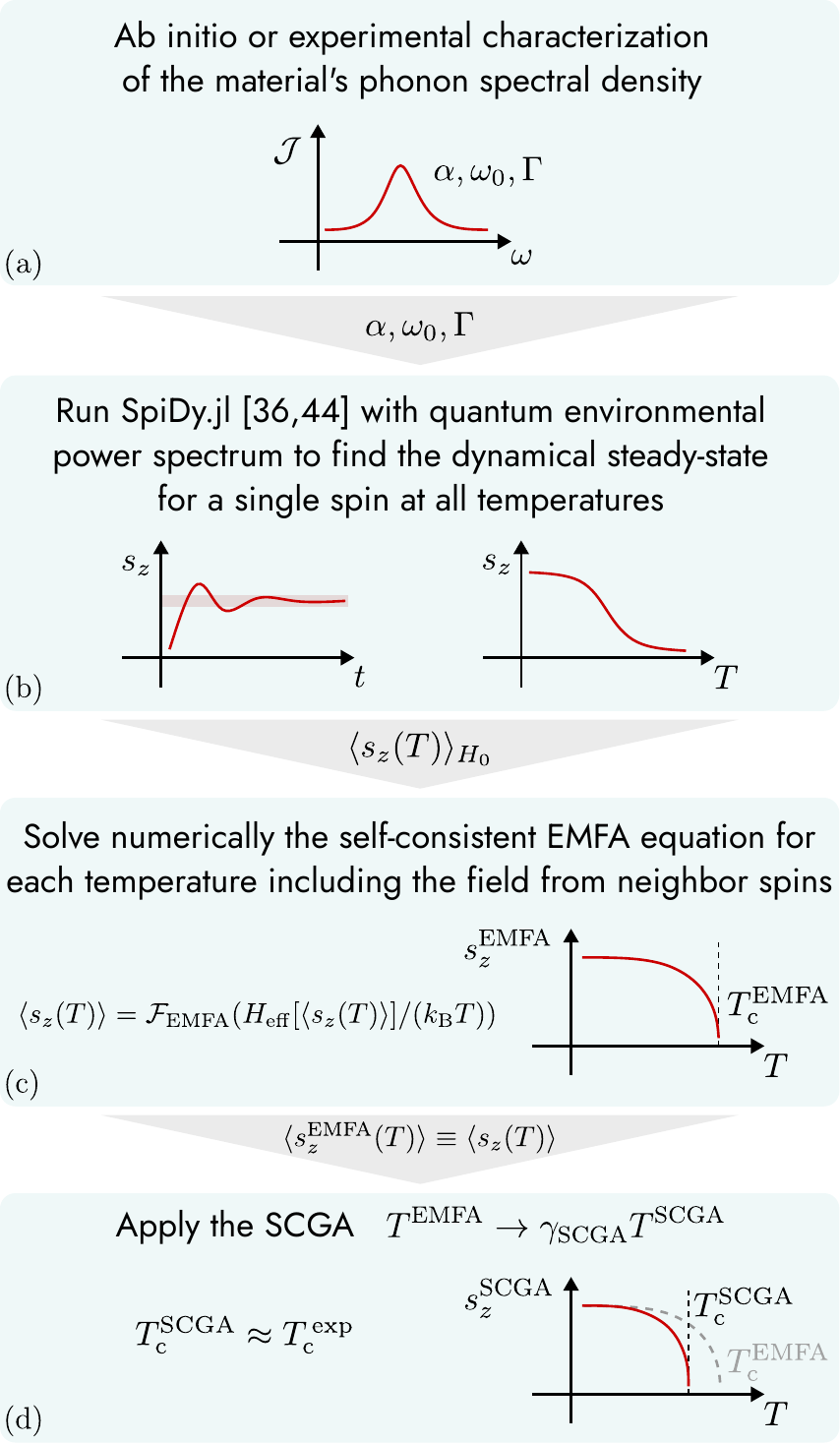}
    \caption{Operational flowchart of the  EMFA approach combined with the self-consistent gaussian approximation (SCGA). 
    \label{fig:flowchart}}
\end{figure}

\begin{enumerate}
    \item[(a)] We parametrize the spectral density $J(\omega)$ in the power spectrum in Eq.~\eqref{eq:semicl} for the external reservoir (in our case phonons) with experimental data according to Ref.~\cite{nemati22} or with {\em ab initio} calculations~\cite{phonopy}.
    \item[(b)] We simulate the dynamics of a single classical spin in a magnetic field $\bm{H}_0$ and interacting with a bath with the quantum power spectrum~\eqref{eq:semicl} for each temperature using the method described in Ref.~\cite{anders20,spidy}. Solving for a long enough time, we find the steady state of the single spin system which aligns along the direction of $\bm{H}_0$. Thereby, assuming $\bm{H}_0=H_0\hat{e}_z$, we find the functional form of $\mathcal{F}_\mathrm{EMFA}(H_{0}/k_BT)=\langle s_z(T)\rangle_{H_{0}}$ in Eq.~\eqref{eq:os--mfa} describing the steady-state of a single spin in an external magnetic field $H_0$ and interacting with a phononic environment at temperature $T$.
    \item[(c)] We numerically solve the self-consistent equation $\langle s_z (T) \rangle =\mathcal{F}_\mathrm{EMFA}(H_{\rm eff}[\langle s_z (T) \rangle]/k_B T)$ using the fixed point method for different $T$, obtaining $s_z^{\rm EMFA}(T)\equiv\langle s_z(T)\rangle$. Note here that the stochastic part of the effective magnetic field (second term of the right-hand-side of Eq.~\eqref{eq:eff_h})
    has already been accounted for in the previous step, therefore $H_\eff [\langle s_z \rangle] =\frac{J}{\gamma} \mathsf{z} S_0 \langle s_z \rangle$. The solution of this equation gives the temperature dependence of the magnetization of the material described by the spin in the Heisenberg model which is additionally coupled to a phonon bath with spectral density $\mathcal{J}(\omega)$. The EMFA method produces a critical temperature, $T_c^{\rm EMFA}$, beyond which the magnetization vanishes $s_z(T>T_c^\mathrm{EMFA})=0$.
\end{enumerate}
Remarkably, for Ni we find 
that the solution of Eq.~\eqref{eq:os--mfa} as a function of the relative temperature $\tau$ (relative to critical temperatures) compares exceptionally well with the Kuz'min curve which approximates the experimental curve (gray). Moreover, our  goal with the EMFA method is to provide a good estimate for the absolute critical temperature $T_c$. Note, that such   prediction is not possible with the heuristic rescaling method discussed in Sec.~\ref{sec:temp_res}. In contrast, the EMFA method we described in this section is suitable to predict critical temperatures. For Ni we find $T_c^{\rm EMFA}=745$~K, which compares reasonably well with the experimental critical temperature of $T_c^{\rm exp} = 632$~K \cite{kuzmin05}.

At this point it is worth remembering that the EMFA method relies on the MFA, which only takes into account  first order fluctuations  around the mean. But it is possible to go to higher orders. To do so we make use of the so called self-consistent Gaussian approximation (SCGA). The SCGA is a field theoretical technique that allows to account for the fluctuations $\mathbf{S}^{(j)}-\langle \mathbf{S}^{(j)}\rangle$ at the second order self-consistently similarly to the random phase approximation (RPA)~\cite{garanin}. Based on Garanin's work \cite{garanin}, Hinzke {\it et al.} showed \cite{hinzke15}, that the effect of including second order fluctuations can be reduced to a linear rescaling of the temperature in the $s_z(T)$ function as obtained by means of the MFA as $T\rightarrow\gamma_{\rm SCGA}T$, where $\gamma_{\rm SCGA}=0.79$.
Therefore, to account for higher order fluctuations and reproduce the absolute temperature dependence of the magnetization we perform a further step.
\begin{enumerate}
    \item[(d)] To account also for second order fluctuations of the spin, we rescale the EMFA temperature as reported in Refs.~\cite{hinzke15,garanin}  as $T\rightarrow\gamma_\mathrm{SCGA}T$ with $\gamma_\mathrm{SCGA}=0.79$. 
\end{enumerate}

\begin{figure}
    \centering
    \includegraphics{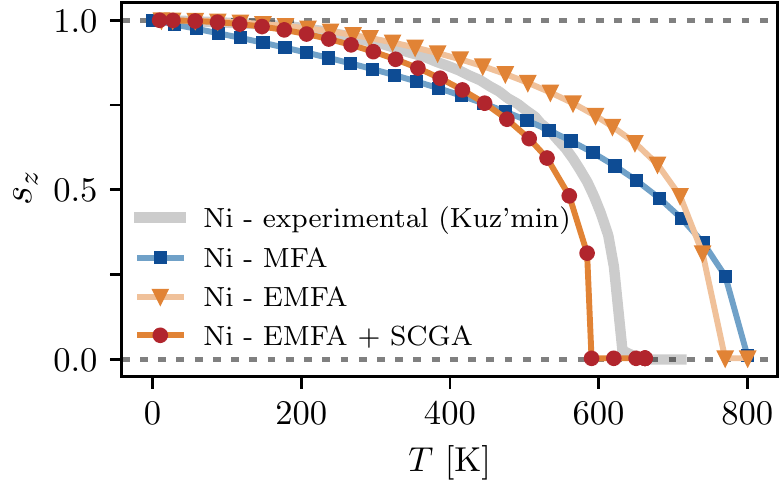}
    \caption{\textbf{Magnetization predicted with the EMFA approach}. EMFA (orange triangles) is compared to the EMFA plus corrections from the self-consistent Gaussian approximation (EMFA+SCGA) (red circles), and the Kuz'min curve (gray solid line). The classical MFA prediction in terms of the physical temperature $T$ is also shown (blue squares) highlighting its qualitative discrepancy with the experiment. Differently to Fig.~\ref{fig:panels}, these magnetization curves are evaluated against the physical temperature in units of Kelvin. If the EMFA and EMFA+SCGA curves were shown over their respective relative temperature $\tau$, they would be the same. We note that both the EMFA and EMFA+SCGA accurately reproduce the low-temperature behavior of the experimental magnetization. Furthermore, EMFA+SCGA gives a relatively accurate prediction of the critical temperature.}
    \label{fig:EMFA}
\end{figure}

In Fig.~\ref{fig:EMFA}, the circles line shows the temperature dependence of the magnetization for nickel as predicted by our approach (EMFA+SCGA). The plot here uses physical temperature in Kelvin ($T[\mathrm{K}]$) instead of relative temperature ($\tau\clqu$). 
We compare the (EMFA+SCGA) approach with bare EMFA (triangles line) and the phenomenological Kuz'min curve~\cite{kuzmin05} (gray solid line), which is known to match experimental behavior. We further plot the bare classical prediction in absolute temperature $s_z^{\rm cl}(T)$ (squares line). As shown in the figure, both EMFA and EMFA+SCGA give a good prediction of the experimental $s_z(T)$ trend as a function of the absolute temperature in K.
In particular, the EMFA gives a prediction for the critical temperature of $T_\mathrm{c}^\mathrm{EMFA}=745\,\mathrm{K}$ while the more accurate EMFA+SCGA provides $T_\mathrm{c}^\mathrm{EMFA+SCGA}=590\,\mathrm{K}$ which is in relatively good agreement with the experimental critical temperature $T_\mathrm{c}^\mathrm{exp}=632\,\mathrm{K}$~\cite{kuzmin05}.

We further note that the EMFA accurately reproduce the low-temperature behavior of the experimental magnetization which is not captured by the classical MFA.
Importantly, this is a direct consequence of the introduction of the environment. Indeed, in the absence of the environment i.e. setting $C_\omega=0$, the EMFA should reduce to the standard MFA, and both classical MFA and MFA+SCGA (without environment nor temperature rescaling) fail to capture the qualitative behavior in this regime.
Importantly, in contrast to the method described in Sec.~\ref{sec:temp_res}, no phenomenological temperature rescaling is employed here.

Finally, we compare our approach to other methods. Our equations of motion, obtained using Eq.~\eqref{eq:open} with the power spectrum in Eq.~\eqref{eq:semicl}, resemble the quantum thermostat approach described in Ref.~\cite{barker19}. In both approaches, the environmental fluctuations follow a quantum power spectrum where the zero-point fluctuations are removed. 
The main difference is that, in our approach, we account for an environment with a structured spectral density (that is with non-Markovian dynamics) instead of white noise. The structured spectral density approach is microscopically motivated and is suited to represent the coupling between spins and environmental degrees of freedom, see Refs.~\cite{nemati22,kresch07}. In addition, we solve the equations of motion for a single spin and use such solution to predict the $M(T)$ curve in the EMFA approximation. In Ref.\cite{barker19}, the authors solve numerically all the coupled equations of motion for all the spins in the system. However, it is worth noting that the approach to including the environment presented here, without the mean-field approximation, could be adapted for ASD simulations. In such case, it would naturally lead to non-Markovian effects in the dynamics, which have been experimentally observed in ultra-fast magnetization dynamics \cite{neeraj21} and have otherwise been introduced ad-hoc into ASD calculations.


\section{Summary and conclusions}
In this article, we presented two different methods to account for quantum corrections in ASD.

Firstly, we showed that, introducing a power-law temperature rescaling, the classical MFA can accurately reproduce the temperature dependence of the quantum MFA magnetization. This power-law rescaling takes the form $\tau\cl=(\tau\qu)^{\alpha(S_0)}$ where $\alpha(S_0)$ is an intrinsic property of the system and depends only on the spin length $S_0$. The rescaling exponents $\alpha(S_0)$ obtained for Ni and Gd appear to be in very good agreement with the values of Ref.~\cite{evans15}, used to reproduce experimental results via dynamical ASD simulations. Here, we give a prediction of the rescaling exponent $\alpha$ as a function of $S_0$, which does not require \emph{a priori} knowledge of the microscopic description of the system and is not a fit of experimental results.
Our approach can be easily combined with ASD to account for quantum effects. This result gives a clear prescription on how to account for quantum effects in ASD simulations. To do this, it is sufficient to calculate the rescaling exponent $\alpha(S_0)$ by using Eq.~\eqref{eq:alpha_S0_fit} and run ASD simulations at a temperature $\tau_{{\rm sim}}=\tau_{{\rm exp}}^{\alpha(S_0)}$.

To avoid the need to ``artificially'' rescale while still accounting for quantum effects, we presented a second method (see Sec.\eqref{sec:open-heisenberg}). In this section, we extended the MFA to the study of a \semi open Heisenberg model in contact with environmental degrees of freedom, whose quantum statistics bring in quantum effects, e.g., that of phonons. The parameters of this quantum environment+MFA approach can be directly calculated by means of {\em ab initio} calculations or extracted from experiments. We further introduced corrections to the (E)MFA by means of the SCGA. We found very good agreement between the magnetization curves of nickel predicted by this method and the experimental results, particularly at low temperatures. The predicted magnetization is furthermore capable of reproducing the absolute experimental $T_\mathrm{c}$ to good accuracy. This suggests that the quantum corrected classical spin model considered here can be used to faithfully predict the dynamics of magnetic materials at all temperatures.


\begin{acknowledgments}

MB and JA gratefully acknowledge funding from EPSRC (EP/R045577/1). JA thanks the Royal Society for support.
SS is supported by a DTP grant from EPSRC (EP/R513210/1).
JA, FC gratefully acknowledge funding from the Foundational Questions Institute Fund (FQXi-IAF19-01).

\end{acknowledgments}

\medskip 

\small{\it For the purpose of open access, the authors have applied a ‘Creative Commons Attribution' (CC BY) licence to any Author Accepted Manuscript version arising from this submission.}


%
\newpage



\vfill


\end{document}